\documentclass[useAMS,usenatbib]{mn2e}
\usepackage{amsmath}
\usepackage{graphicx}
\usepackage{xspace}
\def\araa{ARA\&A}%
\def\apj{ApJ}%
\def\apjs{ApJS}%
\def\mnras{MNRAS}%
\def\prd{Phys.~Rev.~D}%
\def\prl{Phys.~Rev.~Lett.}%
\newcommand{\astroph}[1]{\textsf{astro-ph/}#1}

\def\etal{{\it et al.}\xspace}

\def\ie{{\it i.e.}\xspace}
\def\eg{{\it e.g.}\xspace}

\def\wrt{{\it w.r.t}\xspace}

\def\to                 {\ensuremath{\rightarrow}\xspace}
\newcommand{\C}[1]{\ensuremath{{\cal{C}}_{#1}}\xspace}
\newcommand{\cl}{{\ensuremath{{\cal{C}}_{\ell}}}\xspace}
\newcommand{\clvrai}{\ensuremath{C_{\ell}}\xspace}
\newcommand{\clf}[1]{\ensuremath{{\cal{C}}({#1})}\xspace}
\newcommand{\wmap}{\textsl{WMAP}\xspace}
\newcommand{\planck}{\textsl{Planck}\xspace}
\newcommand{\npar}{\ensuremath{N_\mathrm{par}\xspace}}

\DeclareMathOperator{\sinc}{\sin_\text{c}}
\newcommand{\fmax}{\ensuremath{f_\mathrm{max}}}
\newcommand{\lmax}{\ensuremath{\ell_\mathrm{max}}}
\newcommand{\TF}{\ensuremath{\xrightarrow{FT}}}
\newcommand{\dl}{\ensuremath{\Delta_\ell}\xspace}
\newcommand{\gauss}{\textsl{Gauss}\xspace}
\newcommand{\gsh}{\textsl{Gauss-Shannon}\xspace}
\newcommand{\sh}{\textsl{Shannon}\xspace}
\newcommand{\url}[1]{#1}
\renewcommand{\vec}[1]{\bmath{#1}}
\newcommand{\mat}[1]{\bmath{#1}}
\newcommand{\plmsq}{\ensuremath{{\cal P}_l^m(\cos\theta)}^2}
\newcommand{\plm}{\ensuremath{{\cal P}_l^m(\cos\theta)}}
\newcommand{\pl}{\ensuremath{P_l(\cos\theta)}}

\newcommand{\gm}{{\ensuremath{\Gamma_m(\theta)}}\xspace}
\newcommand{\ct}{{\ensuremath{C(\theta)}\xspace}}
\def\article{article}
\newcommand{\bibi}[3]{\bibitem[\protect\citeauthoryear{#1}{#2}]{#3}}
\newcommand{\hata}{\ensuremath{\hat{\vec{a}}}\xspace}
\newcommand{\hataj}{\ensuremath{\hat{a}_j\xspace}}
\newcommand{\hatVa}{\ensuremath{\hat{\mat{V_a}}\xspace}}

\title[Generic description of CMB power spectra]{Generic description of CMB power spectra}
\author[S. Plaszczynski and F. Couchot]{S. Plaszczynski$ $\thanks{E-mail:
plaszczy@lal.in2p3.fr} and F. Couchot\\
Laboratoire de l'Acc\'el\'erateur Lin\'eaire,\\
     IN2P3-CNRS et Universit\'e de Paris-Sud, BP34- F-91989 ORSAY cedex}
\begin{document}

\date{}

\pagerange{\pageref{firstpage}--\pageref{lastpage}} \pubyear{2003}

\maketitle

\label{firstpage}

\begin{abstract}
Taking advantage of the smoothness of  CMB \cl power spectra, we derive
a simple and model-independent parameterization of their measurement.
It allows to describe completely the spectrum, \ie provide an
estimate of the value and the error for any real $l$ point at the
percent level, down to low $\ell$ multipole.
We provide this parameterization for \wmap first year data and show that the
spectrum is consistent with the smoothness hypothesis.
We also show how such a  parameterization allows to retrieve the \cl spectra
from the measurement of Fourier rings on the sky ($\Gamma_m$) or from
the angular correlation function (\ct).
\end{abstract}


\section{Introduction}

The study of the anisotropies 
of the cosmic microwave background (CMB) temperature has become a
major field in modern cosmology.
For angular scales below $\theta \la 1^\circ$, the \clvrai power
spectrum, defined as the Legendre-transform of the two-point
auto-correlation function \ct, is expected
to present a set of acoustic peaks due to causal physics well
established on the theoretical ground (\eg \citealt{Hu}).

When the first peak began to emerge from the data, it was natural to
characterize its location/amplitude/spread. This was
performed by estimating the maximum in a fixed $\ell$ range, for
instance by fitting a gaussian \citep{Knox} or a
polynomial function \citep{Durrer}. It was further motivated by the fact that
\citet{Hu2} showed that for the first two peaks most of the
cosmological model information was contained in the peaks locations
and relative heights. 

As more peaks became available, in particular in the high $\ell$ region
thanks to interferometer-based experiments, a more complete
\textit{phenomenological} fit was proposed
to determine the peaks locations \citep{Odman} through a sum of
gaussian functions. Adding an oscillatory function has also been
proposed \citep{Douspis} to characterize the existence of the
peaks. While acceptable in the $50\la\ell\la1250$ region (with 5
gaussians) it fails to describe the low $\ell$ Sachs-Wolfe plateau 
and the high part of the spectrum \citep{Odman2}. 
Note that there are no physical reasons for the peaks to have a
gaussian shape.

With the advent of the high precision \wmap results \citep{Bennet} and
the expected huge sensitivity of the future planned \planck satellite 
\footnote{Planck home page:\\ http://astro.estec.esa.nl/SA-general/Projects/Planck/} 
mission, it is time to consider the precise parameterization of the
full \clvrai spectrum over a broad $\ell$ range.

Obviously a modeling through cosmological parameters is not adapted
to describe a purely experimental spectrum.
We propose to take advantage of the expected \textit{smoothness} of
the 
\clvrai power spectrum, which comes from a combined effect  of the continuity
of the Fourier $P(k)$ spectrum in the Standard Model, and the use of
spherical Bessel functions to project it onto the $\ell$ space (\eg \citealt{Bartlett}).

This smoothness property has already been exploited for fitting the
\cl spectrum with splines (\eg \citealt{Oh}).
Here we will rather work in the light of the Fourier decomposition in
particular by revisiting the sampling theorem. This will allow to
provide a very simple description of any \clvrai spectrum as a
\textit{function} of \textit{real} $\ell$ values.
Therefore our goal is twofold:
\begin{enumerate}
\item obtain a reduced number of parameters to describe a CMB power spectrum
\item provide an interpolation for any (real) $\ell$ value, \ie a
  central value and a (Gaussian) error.
\end{enumerate}

\section[]{Describing $\bmath{\cl}$ spectra}
\label{sec:method}

\paragraph*{The sampling theorem}

A CMB power spectrum $\cl\equiv \ell(\ell+1)/2\pi~ \clvrai$ can be
considered as the sampling for integer values of a smooth continuous
signal $\clf{l}$, where $l$ is real. 
The signal is band-limited so that its Fourier transform can be
neglected for $|f|>\fmax$.
The sampling theorem states that for 
any rate (\dl) above the critical frequency:
\begin{equation}
  \label{eq:shannon}
\dfrac{1}{\dl} = 2\fmax
\end{equation}
the real space complete signal can be recovered through:
\begin{equation}
\label{eq:interp}
  \clf{l}=\sum_j \C{j} \sinc{\left(\pi\frac{l-l_j}{\dl}\right)}
\end{equation}
where $l_j$ are the sample positions.

\begin{figure}
\centering
\includegraphics[width=85mm]{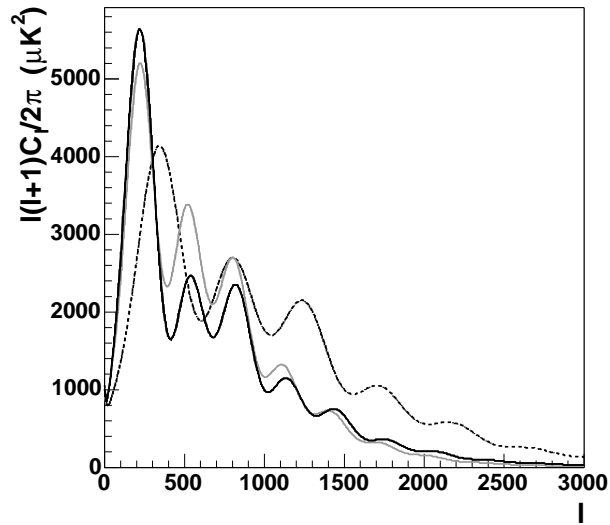}
\vspace{-1cm}
  \caption{Three different CMB TT power spectra (up to $\lmax=3000$),
    used as illustrations for the method.     
    The dark full line corresponds to \wmap data best fit model.}
  \label{fig:models}
\end{figure}

It appears that ``reasonable" CMB power spectra are (mostly) band-limited
to low frequencies. We
illustrate that feature by choosing three representative spectra
(Figure \ref{fig:models}): 
one is the ``\wmap best fit model" ($\Lambda CDM$ of \cite{Spergel}), a second one  has different peak proportions, and a third one has shifted peak locations.

\begin{figure}
\includegraphics[width=85mm]{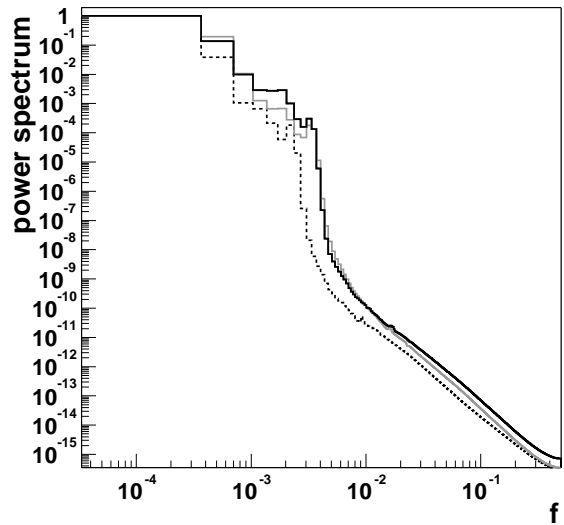}
\vspace{-1cm}
  \caption{Power spectral densities (periodogram) of the three \cl
    spectra presented on Figure \ref{fig:models}.}
  \label{fig:fft}
\end{figure}

We then show their power spectral densities (estimated through a periodogram) on Figure \ref{fig:fft}.

In each case, most of the power lies below $\fmax\simeq 0.005$ which
corresponds to a minimum sampling rate of :
\begin{equation}
  \Delta_\text{min} \simeq 100
\end{equation}

This indicates that only 1 parameter on 100 $\cl$ values is necessary to
reconstruct the whole spectrum.

\paragraph*{Fitting $\bmath{\cl}$ spectra}

When using the sampling rate $\dl=\Delta_\text{min}$, 
some part of the Fourier spectrum is nonetheless removed, and the
Shannon interpolation
given by \eqref{eq:interp} is no longer exact. We therefore modify it to
a linear parameterization :
\begin{equation}
  \label{eq:base}
  \clf{l} =\sum_{j=1}^{\npar} a_j u\left(\dfrac{l-l_j}{\dl}\right)
\end{equation}

where:
\begin{description}
\item - the $\{a_j\}_{j=1,...,\npar} $ represent a set of parameters
  to be estimated on the data through least square minimization.
\item - $u$ denotes the ``basis" function and will be discussed more in
  details in the next part.
\item - the $\{l_j\}$ represent a grid of \textit{fixed} points:
  $l_j=l_0+j \dl$. Note that the first point ($l_0$) 
  can be located anywhere. In practice one fixes it to the first
  measured multipole.
\end{description}

Eq. \eqref{eq:base} denotes a decomposition of the \clf{l} function over
the non-orthogonal basis $u_j(l)= u[(l-l_j)/\dl]$.

Given a set of ``data" points, \ie a set of \cl measurements and their
covariance matrix, the parameters  $\{a_j\}$  and their covariance
matrix are determined through a simple linear least square fit. 
Then one has an estimate of the $\clf{l}$ function for any real 
$\ell$ value together with its associated Gaussian error through
standard error propagation (see section \ref{sec:wmap} for an example).

\paragraph*{Bases}

The sampling theorem uses the basis (denoted as \sh):
\begin{equation}
  \label{eq:sh}
  u(x)=\sinc{(\pi x)}
\end{equation}
It has the the nice property of being an exact interpolation
on each point of the grid:
\begin{equation}
u(\dfrac{l_i-l_j}{\dl})=\delta_{ij}
\end{equation}
so that the estimated parameters (\hataj) satisfy:
\begin{equation}
\label{eq:exact}
  \clf{l_j}=\hataj
\end{equation}
and can be estimated ``by eye" on a (perfect) \clf{l} curve.

However the \sh basis is not strongly local in real space, involving long
range correlations in the parameters covariance matrix.

To modify this behavior but still keeping the exact interpolation
property, we multiply it by a Gaussian
function and call this new basis \gsh:
\begin{equation}
  \label{eq:gsh}
  u(x)=\sinc{(\pi x)}\times  \exp{(\dfrac{-x^2}{2\sigma^2})}
\end{equation}

With respect to the Shannon basis, the off-diagonal terms of the
parameters covariance matrix will be reduced.
Although in principle one has the freedom to adjust the $\sigma$ of the
Gaussian, it is more interesting to use ``large" values, since in that
case the bulk part of the Fourier spectrum is un-filtered : one just
cuts off frequencies near $\fmax$ and the Shannon properties are (mostly)
preserved. In the following we will therefore investigate the
$\sigma=3$ case in Eq. \eqref{eq:gsh}.

For the sake of comparison, we will also investigate a very different 
basis provided by the
simple Gaussian function\footnote{Note that this
  approach is different from the one proposed by \cite{Odman} where
  \textit{all} the parameters of the gaussians are fitted.} (\gauss basis):
\begin{equation}
u(x)=\exp(\dfrac{-x^2}{2\sigma^2})
\end{equation}
It does not have the exact interpolation property of
Eq. \eqref{eq:exact} since it
is always positive. Therefore the estimated parameters \hataj
will now be very different from the $\clf{l_j}$  function values, and the off-diagonal terms of the covariance matrix will be important.
The Gaussian function is however the most rapidly decaying in both real
and Fourier spaces \citep{hardy} and is therefore of great interest.
If we want to perform a fit to Eq. \eqref{eq:base}, we
do not have the freedom in the choice of $\sigma$. Indeed, since:
\[
u(\frac{l}{\dl}) \TF \dl U(\dl f) \propto \exp(-2 \pi^2
(\dl\sigma)^2 f^2)
\]
by changing $\sigma$ we just define a new effective sampling rate. If
we wish to keep the Shannon critical frequency of
Eq. \eqref{eq:shannon} we are lead to use $\sigma=1$ to avoid
aliasing or over-sampling.

\paragraph*{Edge effects}
\label{sec:side}
The Shannon theorem, on which is based the fit, is defined on an
infinite support. By restricting the data to some measured $\ell$ range,
a high frequency ringing appears near the boundaries. 
To reduce this  phenomenon one can
extend the grid beyond the data limits. Given
the short range of the functions basis used in Eq.\eqref{eq:base} , 
only one or two points can be added but this is sufficient in most of
the cases. Note that we do not add any fake data, but just increase the
grid position and number of parameters to be determined from the fit.

\begin{figure}
\centering
\includegraphics[width=90mm]{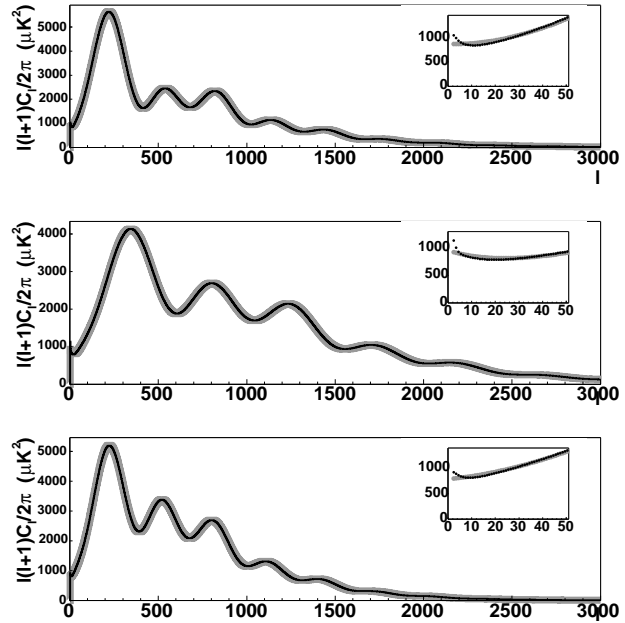}
  \caption{ 
Superposition of the three input \cl spectra (dark thin line) from 
Figure \ref{fig:models}, and the
parameterizations described in the text (thick grey) .
 The \gsh basis has been used, with a sampling rate of $\dl=100$
and 2
extra-points on the low side and one on the high side (therefore 34
parameters have been fitted). The insets show a
close-up of the low $\ell$ region.
For the sake of visualization, the thick grey
curves have been enlarged.
\label{fig:interpres}}
\end{figure}

\paragraph*{Results}
To test the precision of the method, we fit the three \cl spectra of
Figure \ref{fig:models}. We use
the \gsh basis with a grid of 34 points ($\sl=100$ and 2 points added on the
low-side $l<2$ and 1 on the high side $l>3000$).
Since the input data has no error, their covariance matrix is defined as
identity.
Figure \ref{fig:interpres} shows the comparison of the parameterization
and the genuine spectrum. A close-up of the low $\ell$ region is also displayed
The agreement is very satisfactory up to very low $\ell$ values.
The first few bins are not perfectly described because they
introduce a high frequency component that is cutoff by the sampling rate of
$\dl=100$. This can be
accounted for by a second step decomposition of the residuals as in a
wavelet analysis.
For the sake of simplicity and since 
experimental error bars are high in this region, we choose not to
correct for this effect in the following.
The upper part of the spectrum is correctly described.

\begin{figure}
\centering
\includegraphics[width=90mm]{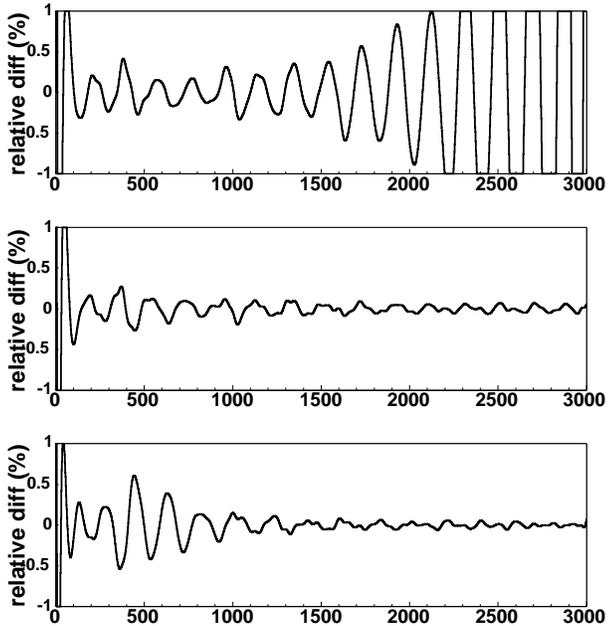}
\caption{Relative agreement (in \%) between the input \cl (\wmap best
  fit model) and the function obtained by fitting 34 parameters on the
  \sh (upper plot), \gsh (middle) and \gauss (bottom) bases} 
  \label{fig:interprel}
\end{figure}

To compare the interest of using different bases,
Figure \ref{fig:interprel} presents the relative agreement between the
fitted values and the input \wmap best-fit model for the three bases: 
\sh, \gsh,\gauss. For the modified bases
(\gsh,\gauss) , the agreement is better than 1\% in the range
$\ell>30$. For low $\ell$ values the agreement decreases up to 
$\simeq 15\%$ for $\ell=2$.

Both \gauss and \gsh give very good results, \gauss being 
slightly more precise for the models we use. Recall however that the
\gauss basis (unlike \gsh) provides parameters with values far from
the grid points and highly correlated. We will therefore prefer the
\gsh basis in the following.

\paragraph*{Derived results}

Once the parameters \hataj have been determined, including their
covariance matrix \hatVa, one can derive easily many characteristics
of the spectrum.

\cl values can be estimated for any (real) $l$ value. 
Using again the notation $u_j(l)= u[(l-l_j)/\dl]$: 

\begin{equation}
  \begin{split}
     \hat{\cl}& =\sum_{j=1}^{\npar} \hataj u_j(l)\\
     \hat\sigma_l^2&= \vec{U} \hatVa \vec{U}^T
   \end{split}
\end{equation}
$\vec{U}$ being the vector of components $U_j=u_j(l)$.

The peak (and dip) locations can be determined by setting the derivatives of
Eq. \eqref{eq:base} to zero and finding (numerically) the roots.

One can compute also binned values. 
For a given binning $(b_1,b_2,..,b_N)$, assuming a weighting scheme
$w_i^b$ inside the bin $b$:
\begin{equation}
     \hat{\cl}^b =1/W_b\sum_{l\in b} w_i^b \hat{\cl}
\end{equation}
with the normalization $W_b=\sum_{l\in b} w_i^b$.

The covariance matrix of the binned estimate is obtained again
through error propagation $\mat{B}=\mat{D}\hatVa \mat{D}^T$, $\mat{D}$ being
a ($N\times\npar$) matrix defined by:
\begin{equation}
     D_{bj}=1/W_b\sum_{l\in b} w_i^b u_j(l)
\end{equation}

\section[]{Application to \wmap data}
\label{sec:wmap}

We now turn on to real data, by describing the \wmap TT angular power spectrum
\footnote{first year data \textsl{v1p1} version, as available from
\url{http://lambda.gsfc.nasa.gov/product/map}}.

The data consist of a sample of N=899 measurements for integer $\ell$
values in the [2,900] range, and of the weight matrix (inverse
covariance) $\mat{W}$ defined as the Fisher matrix for the ML estimate. 

\begin{figure}
\includegraphics[width=95mm]{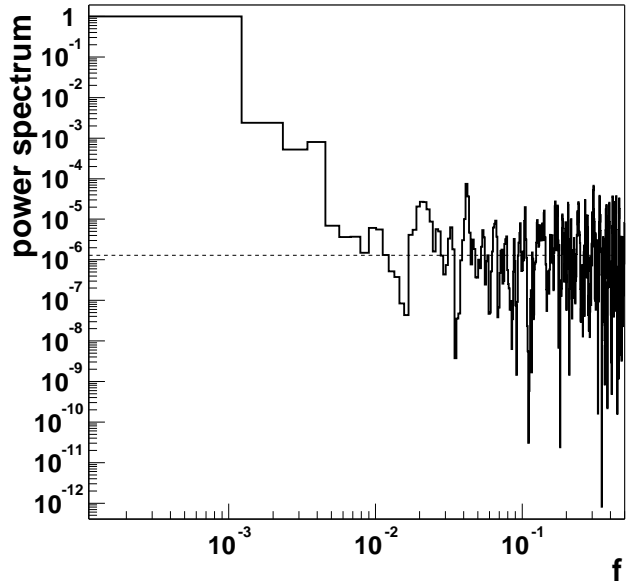}
\vspace{-1cm}
\caption{Power spectrum estimated through the periodogram of \wmap
  data. The dashed line corresponds to the estimate of the mean level
  of noise.\label{fig:fftwmap}}
\end{figure}

We begin with a check of our \cl smoothness hypothesis, 
showing the periodogram of the data on Figure
\ref{fig:fftwmap}. Unlike Figure \ref{fig:fft},
this one has (statistical) noise, described by the covariance matrix of the data ($\mat{C}$).
We estimate the mean level of noise power in each Fourier mode $k$ by:
\begin{equation}
  \label{eq:noise}
  <\lvert n_k \rvert^2>=\sum_{l=0}^{N-1}\sum_{m=0}^{N-1}
  \exp[j\dfrac{2\pi}{N}(l-m)k] \mat{C}_{lm}
\end{equation}
and show it on the Figure too.

The (noise subtracted) 
signal does not show high frequency components and we keep
the cut $\fmax=0.005$ determined previously, leading to  the sampling
rate $\dl=100$.
We then choose the \gsh basis and put 10 $l_j$ knots between 2 and 900.
We add two extra-points on the low $\ell$ side
and one on the high side.
We are therefore left with determining 13 parameters from a set of N=899
input data.

The (linear) least square estimate reads:
\begin{equation}
  \label{eq:fit}
  \begin{split}
  \hata&=(\mat{A W A})^{-1}\mat{A^T W} \vec{c} \\
  \hatVa&=(\mat{A W A})^{-1}
  \end{split}
\end{equation}
with
\begin{equation}
  \mat{A}_{ij}=u(\dfrac{\ell_i-\ell_j}{\dl})
\end{equation}
and $\vec{c}$ is the input data values vector.

There is a subtlety however.
The Fisher matrix (or ``curvature" as defined in \citet{Verde}) depends
actually on the \textit{true} \cl distribution through the cosmic
variance.  We need a means to incorporate it in our fits, in
particular for the low $\ell$ range. This is
performed through the following iterative procedure:
we start from the \wmap best \cl estimates (as true values) 
to obtain the weight matrix, and perform the fit.
We then recompute the weight from the Fisher matrix using this time
the \textit{fitted} \cl values, and redo the fit. 
We pursue the iteration until the $\chi^2$ gets stable. 

For this data set, the fit is stable after 3 iterations.

\begin{figure}
\includegraphics[width=95mm]{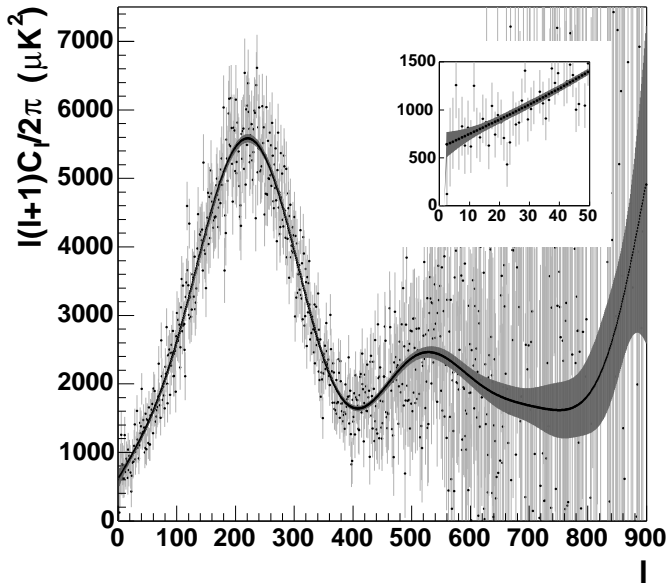}
\vspace{-1cm}
\caption{ Result of the parameterization performed on \wmap data using the
  \gsh basis. Light grey points corresponds to \wmap data for
  each integer $\ell$. Their error is computed from the diagonal
  elements of the inverse of the Fisher matrix as obtained in the last
  step of the iterative procedure described in the text. The dark
  line corresponds to our best estimate for each integer $\ell$ and
  the errors (dark grey area) are computed from the parameters
  covariance matrix through standard error propagation. A close-up of
  the low $\ell$ region is also shown.
  \label{fig:fitwmap}}
\end{figure}

Figure \ref{fig:fitwmap} shows the result of the last fit.

\begin{figure}
\includegraphics[width=80mm]{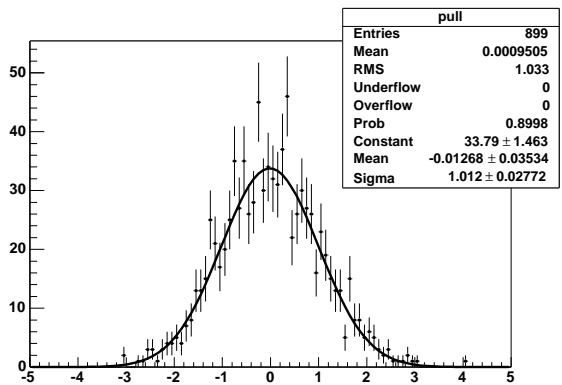}
\caption{ ``Pull" distribution of \wmap data \wrt to our best fit
  estimate. Also superimposed is the fit to a gaussian function. The
  parameters of the histogram and the result of the fit are given in
  the box.\label{fig:pull}}
\end{figure}

The $\chi^2$ per degree of freedom is excellent: $959/886$.
Figure \ref{fig:pull} shows the ``pull" distribution that we define as
the data value minus the estimated value divided by the data error.
Here again the data errors are obtained from the diagonal elements of the
inverse of the Fisher matrix. The distribution is compatible with a
normal one. We checked that this feature is valid over the whole
$\ell$ range.
These results indicate that
\textit{the data are consistent with the assumption of a smooth power spectrum}. 

As is clear from Figure \ref{fig:fitwmap} our fitted
function (which includes the effect of the cosmic variance) has a
positive slope in the low $\ell$ region.

\begin{figure}
\includegraphics[width=90mm]{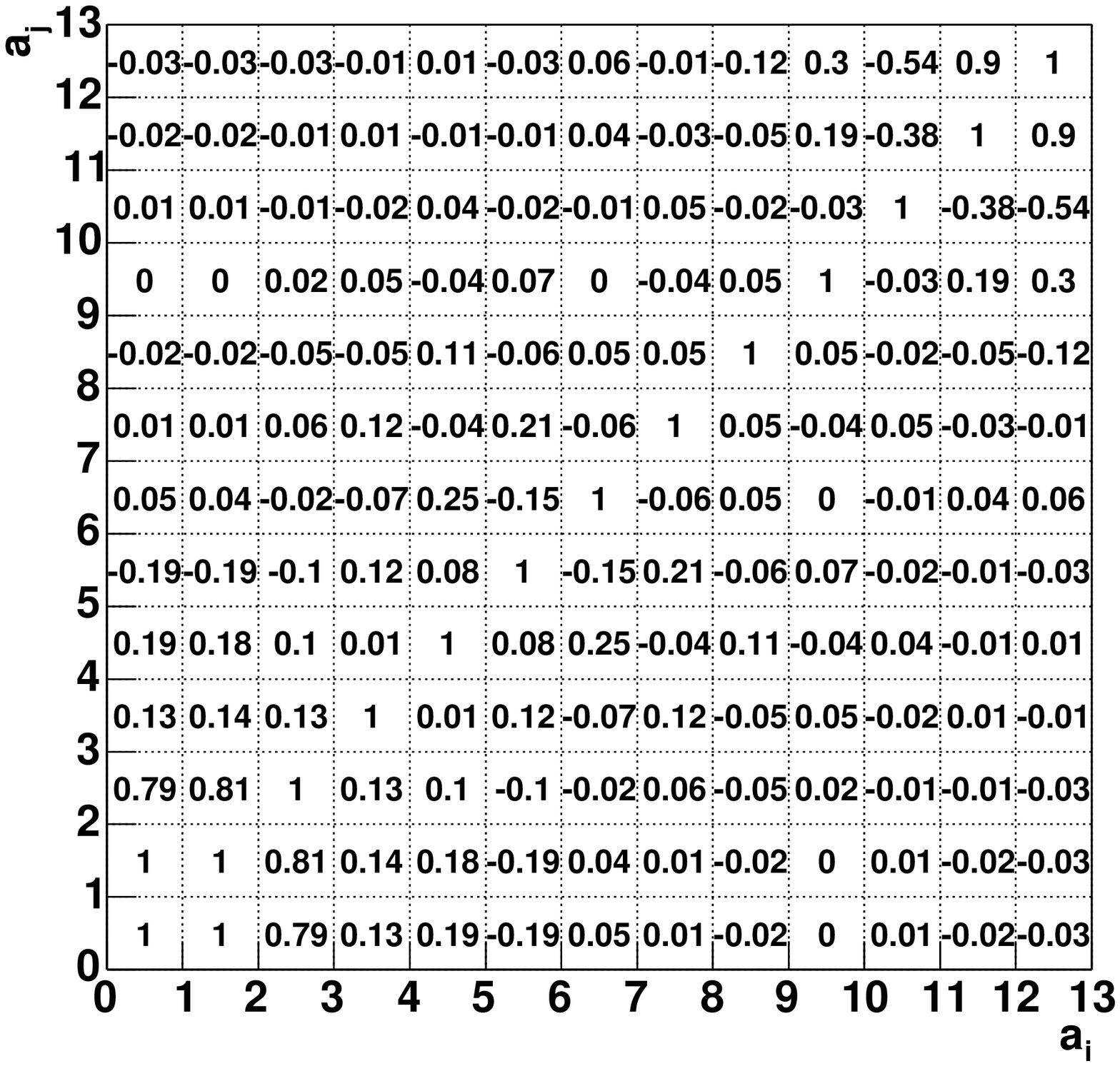}
\vspace{-.5cm}
\caption{Correlation matrix of the parameters determined from the fit
  of \wmap data on the \gsh basis. \label{fig:corr}}
\end{figure}

Finally we provide the parameterization obtained in Table \ref{tab:param}
(parameters) and
Figure \ref{fig:corr} (correlation matrix). 
It represents the most probable estimate of the band limited \cl
spectrum and one can 
derive several secondary results as explained in
section \ref{sec:method}. 
Note how the parameter values can be determined directly from
Figure \ref{fig:fitwmap}, as expected for
the \gsh basis. The off-diagonal terms of the parameters matrix
have also been reduced \wrt a simple \sh based fit.

The three bases give however similar results for the final fitted function.

We emphasize once more that this description is 
independent from any cosmological model.

\begin{table}
  \centering
  \begin{tabular}{|c|c|c|}
\hline
parameter index $j$  &   grid position $l_j$  &  estimated parameters \hataj  \\
\hline
0&      $-197.6$ &       $15.1\pm16087.1$ \\
1&      $-97.8$ &        $-415.9\pm8077.1$ \\
2&      $2.0$ &  $639.9\pm129.8$ \\
3&      $101.8$ &        $2679.8\pm36.4$ \\
4&      $201.6$ &        $5472.4\pm58.6$ \\
5&      $301.3$ &        $3892.7\pm43.6$ \\
6&      $401.1$ &        $1649.7\pm37.4$ \\
7&      $500.9$ &        $2379.8\pm66.6$ \\
8&      $600.7$ &        $2098.0\pm124.6$ \\
9&      $700.4$ &        $1683.9\pm252.8$ \\
10&     $800.2$ &        $1831.2\pm559.8$ \\
11&     $900.0$ &        $4953.8\pm2384.6$ \\
12&     $999.8$ &        $6123.9\pm7475.2$ \\
\hline
  \end{tabular}
  \caption{Value of the parameters (and grid spacing) obtained from
    the fit to \wmap data on the \gsh basis.}
  \label{tab:param}
\end{table}

\section{Other power spectra}

\subsection{Fourier transform of rings \gm}
\label{sec:gm}

For a circular scanning strategy, the Fourier decomposition of circles is an
interesting analysis method , since it does not use map
projection and allows to follow in time the detector response,
avoiding hypotheses such as noise stationarity.

The power spectrum is simply related to the underlying
\clvrai through \citep{Delab}
\begin{equation}
  \label{eq:gmdef}
  \gm=\sum_{\ell\ge m} \plmsq B_l^2 \clvrai
\end{equation}
where
\begin{enumerate}
\item $\theta$ is the opening angle on the sky (colatitude)
\item ${\cal P}_l^m$ denote the (normalized) associated Legendre polynomials
\item $B_l$ is the beam transfer function, depending just on $\ell$
  for a Gaussian symmetric beam. 
\end{enumerate}

The inverse process of retrieving \clvrai values from a set of
measured \gm is delicate due to the fact that $\plm
\to 0$ when $\ell > m/\sin\theta$. \citet{Ansari} have shown
formally how a Fourier scaling in the flat sky limit 
allows to invert the problem up to very
low $\ell$ values ($\ell \ga 3$ for ($\theta=40^\circ$)).
The key point in their approach is to interpolate the \gm spectrum to
non-integer $m$ values ($m/\sin\theta$) which can be done with the
present method, since the  $m\gm$ spectrum is smooth. 
It is simpler however to stay in the $\ell$ space and use the previous
results.

We describe again the \clf{l} function in terms of
eq. \eqref{eq:base}. By combining with Eq. \eqref{eq:gmdef} :
\begin{equation}
  \label{eq:gmfit}
  \gm=\sum_j a_j \left[ \sum_{\ell \ge m} \dfrac{2\pi B_\ell^2}{\ell(\ell+1)} \plmsq
  u(\dfrac{\ell -l_j}{\dl})\right]
\end{equation}

which is still linear in the $a_j$.

The $a_j$ parameters can therefore be fitted directly on the \gm~ data using
the above new basis, and the
\cl spectrum is still obtained from Eq.\eqref{eq:base}.

Note that this approach allows to combine any number of
detectors with different opening angles and transfer functions.

\subsection{Angular power spectrum \ct}

Using again the decomposition (Eq. \eqref{eq:base}) , the two-point
angular correlation function can be expressed linearly in terms of the
$a_j$:
\begin{equation}
  \label{eq:ct}
  \ct=\sum_j a_j \left[ \sum_{\ell} \dfrac{2\ell+1}{2\ell(\ell+1)} \pl
  u(\dfrac{\ell -l_j}{\dl})\right] 
\end{equation}
The same approach than in the previous part is therefore possible.
Although equivalent to the usual methods, it allows to drop the $a_{lm}$
computations and adjust directly the \cl distribution, even on partial
or masked sky data.

\section{Conclusion}

In this \article, we take advantage of the smoothness of the \cl power
spectra to decompose the signal in Fourier space and apply an improved
version of the
sampling theorem. We obtain an accurate parameterization of the
spectra, that is independent from cosmological models, as a function
$\clf{l}$  for real $l$, with few parameters: for ``reasonable" CMB
models we find that sampling the \cl spectrum with one point on 100 is
sufficient to retrieve the signal on any $\ell$ range at the percent
level, down to very low $\ell$ values. 
We also show how this kind of description can be applied to other CMB
power spectra as the Fourier spectrum of rings ($\Gamma_m(\theta)$) or
the angular correlation function ($C(\theta)$).

We apply the method to \wmap first year data and provide
the complete parameterization of the \cl spectrum. We show that the
data is consistent with the expected smoothness of the spectrum.

Such a parameterization is richer than peaks determination (which can
obviously be derived from it). It also allows the combination of
various experiments and a $\chi^2$ test of the compatibility between
them based on the assumed smoothness of the spectrum.

Finally it can allow to compress the data (by a factor $\simeq 100$)
for the storage of cosmological models used in large databases. 

The method can be applied to any band-limited function to be adjusted
on data.

\section*{ACKNOWLEDGMENTS}

We are pleased to thank Nabila Aghanim for informations on
previous existing works, and Sophie Henrot-Versill\'e, Jacques
Ha\"{\i}ssinski and Alexandre Bourrachot for fruitful discussions and 
pertinent reading of the document.


\bsp

\label{lastpage}

\end{document}